\documentclass[aps,prl,twocolumn,showpacs]{revtex4-1}

\usepackage{amsmath}
\usepackage{amssymb}
\usepackage{bm}
\usepackage{textcomp}
\usepackage{graphicx}
\usepackage{bbold}

\begin{document}

\title{Entanglement Purification with the Exchange Interaction}


\author{Adrian Auer}
\email{adrian.auer@uni-konstanz.de}
\author{Guido Burkard}
\affiliation{Department of Physics, University of Konstanz, D-78457 Konstanz, Germany}

\begin{abstract}
Entanglement purification allows the creation of qubit pairs of arbitrarily high fidelity with respect to a maximally entangled state, 
starting from a larger number of low-fidelity pairs.  
Purification requires quantum memory, a role for which electron spins are well suited.
However, using existing recurrence protocols involving symmetric local two-qubit operations
for spin qubits turns out to be rather unpractical. 
We present an efficient purification protocol requiring only a single pulsed Heisenberg- or XY-type exchange interaction between two qubit pairs. 
In contrast to known protocols, we allow for asymmetric bilateral two-qubit operations where the two communication parties operate differently on their qubits.
In the optimal version of our protocol, the local two-qubit interactions in the case of Heisenberg exchange correspond to the $\sqrt{\textsc{swap}}$ gate and its inverse. 
\end{abstract}


\maketitle

\newcommand{\bra}[1]{\langle #1|}
\newcommand{\ket}[1]{|#1\rangle}
\newcommand{\braket}[2]{\langle #1|#2\rangle}

\textit{Introduction.}
Many quantum communication (QC) protocols \cite{bouwmeester2000physics} rely on the faultless distribution of maximally entangled two-qubit states. Well-known examples are quantum teleportation \cite{PhysRevLett.70.1895} and quantum dense coding \cite{PhysRevLett.69.2881}, while quantum key distribution is possible without \cite{BB84} or with entanglement \cite{PhysRevLett.67.661}. Due to imperfect sources and the inevitable interaction of the entangled particles with their environment, however, the degree of entanglement decreases, and the security or feasibility of the mentioned QC protocols cannot be guaranteed. In the realistic scenario of noisy channels the entanglement can be restored using the method of \textit{entanglement purification}, also known as \textit{quantum privacy amplification} \cite{PhysRevLett.76.722,PhysRevLett.77.2818,0034-4885-70-8-R03}.

The idea is to use two or more imperfectly entangled qubit pairs to purify one of them with respect to a maximally entangled state. The principle of entanglement purification is specifically harnessed as a building block of quantum repeaters \cite{PhysRevLett.81.5932,PhysRevA.59.169}. In large-scale QC networks \cite{Kimble:2008}, long-distance entanglement between stationary qubits, acting as quantum memory \cite{Simon:2010}, has to be established, and photons are typically used to transfer quantum information between them. Due to lossy communication channels, e.g.~optical fibers, photons can be lost or decohere, which impedes the entanglement production between remote nodes. The working principle of the quantum repeater established in \cite{PhysRevLett.81.5932,PhysRevA.59.169} is to divide the distance between the network nodes into smaller segments, create entangled states between them, purify these states individually, and finally connect them via entanglement swapping \cite{PhysRevLett.71.4287}.

A promising candidate for the realization of the stationary qubits are electron spins, e.g.~in semiconductor quantum dots (QDs) \cite{PhysRevA.57.120}. Due to their long coherence times of several $\mu$s \cite{PhysRevLett.100.236802} and their complete controllability by electrical \cite{RevModPhys.79.1217} or optical \cite{Berezovsky18042008} means, spin qubits have considerable potential for quantum computation or as quantum memories. However, the original purification proposals \cite{PhysRevLett.76.722,PhysRevLett.77.2818} make use of a bilaterally applied controlled-\textsc{not} (\textsc{cnot}) operation. In the case of spin qubits in tunnel-coupled QDs, the interaction is of Heisenberg-exchange type, which needs to be supplemented with single-spin rotations  to produce a \textsc{cnot} gate \cite{PhysRevA.57.120}. It would therefore be advantageous to perform entanglement purification directly using the exchange interaction. In this paper, we present a simple purification protocol based solely on the one-time activation of a Heisenberg interaction leading to the $\sqrt{\textsc{swap}}$ gate. Following the same approach, we also find a similar result for qubits coupled via a XY-type interaction, which happens to be the interaction between superconducting qubits \cite{RevModPhys.73.357} as well as between optically coupled spin qubits \cite{PhysRevLett.83.4204}.

Earlier works have demonstrated purification schemes for spin qubits e.g.~by replacing the \textsc{cnot} in the bilateral operation by the gate sequence that uses two-qubit gates directly generated from the interaction Hamiltonian \cite{PhysRevA.78.062313}, requiring additional single-qubit operations. Other procedures use three input pairs \cite{PhysRevA.78.022312,PhysRevA.84.042303,PhysRevA.86.052312} or specifically work for two-spin singlet-triplet qubits \cite{PhysRevLett.94.236803}. Our proposal works with two input pairs of spin 1/2 qubits and only requires a single two-qubit interaction. In comparison with existing protocols, we achieve an advantage by allowing for \textit{different} two-qubit manipulations locally in the bilateral operation.

\textit{Entanglement Purification.}
A basis of the two-qubit Hilbert space is given by the maximally entangled Bell states
$\ket{\Phi^\pm} =
\frac{1}{\sqrt{2}}( \ket{00} \pm \ket{11})$ and
$\ket{\Psi^\pm} = \frac{1}{\sqrt{2}}(\ket{01} \pm \ket{10})$,
where $\{ \ket{0},\ket{1} \}$ is the single-qubit logical basis of the sharing parties, conventionally named Alice and Bob, and we call the overlap of an arbitrary state $\rho$ with the desired entangled state $\ket{\Phi^+}$ its \textit{fidelity} 
$F\equiv \bra{\Phi^+} \rho \ket{\Phi^+}$.
Recurrence protocols work on two or more qubit pairs of low fidelity as input to create a single qubit pair with higher fidelity as output using only local unitary operations, measurements, and two-way communication of the measurement results via a classical channel. Having initially many copies of the low-fidelity pairs and running the purification protocol iteratively on the output pairs with higher fidelity, one can achieve fidelities arbitrarily close to $F=1$ and thus obtain a maximally entangled state.

\begin{figure}
\includegraphics[width=0.5\textwidth]{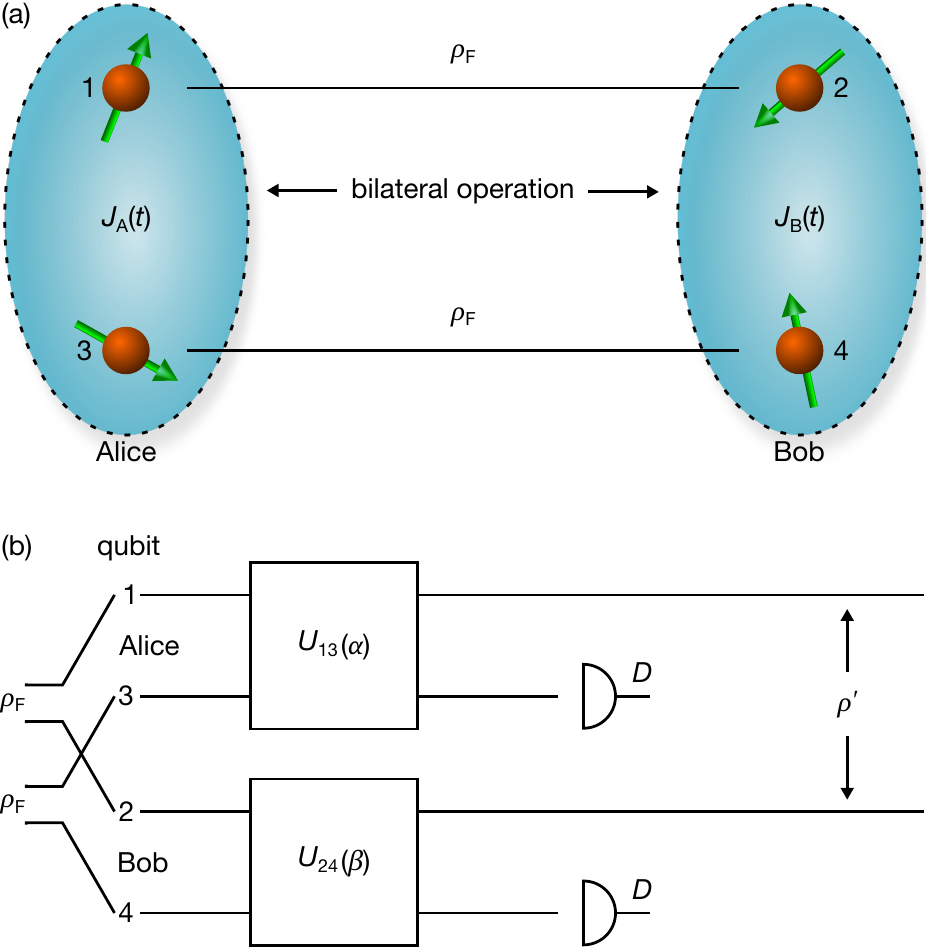}
\caption{\label{fig:circuit} (a) Alice and Bob share two imperfectly entangled qubit pairs $\rho_F$ and locally apply different unitary operations generated by the exchange couplings $J_\text{A}$ and $J_\text{B}$. (b) Circuit diagram of the protocol explained in the main text, where unitary operations with different pulse areas $\alpha$ and $\beta$ are applied. After the detection ($D$) of qubits 3 and 4, Alice and Bob are left with the two-qubit state $\rho'$.}
\end{figure}

The protocol of Bennett \textit{et al.} \cite{PhysRevLett.76.722} (\textsc{bbpssw}) requires two copies of the state $\rho_F$ (Fig.~\ref{fig:circuit}),
\begin{eqnarray}
\rho_F
&&=
F \ket{\Phi^+} \bra{\Phi^+} 
\nonumber \\
&&+ \frac{1-F}{3}
\Big(
\ket{\Phi^-} \bra{\Phi^-} + \ket{\Psi^+} \bra{\Psi^+} +  \ket{\Psi^-} \bra{\Psi^-}
\Big),
\end{eqnarray}
which can be generated from an arbitrary two-qubit state having overlap $F$ with the rotationally invariant singlet state $\ket{\Psi^-}$ by a \textit{twirl} operation \cite{PhysRevLett.76.722,PhysRevA.54.3824} that retains the singlet component, equalizes the triplet components and removes all off-diagonal elements, thus creating a so-called Werner state \cite{PhysRevA.40.4277}, followed by a $\pi$ rotation about the $y$ axis on the Bloch sphere by one of the sharing parties, hence interchanging the $\ket{\Psi^-}$ and $\ket{\Phi^+}$ components.
Next, Alice and Bob each perform a \textsc{cnot} gate between the two qubits they hold, respectively, where the qubits of the first (second) pair serve as source (target) bit and the target bit is flipped if the source qubit is in state $\ket{1}$, i.e.,
$\ket{00} \mapsto \ket{00}$, $\ket{01} \mapsto \ket{01}$,
$\ket{10} \mapsto \ket{11}$, and$\ket{11} \mapsto \ket{10}$.
After this bilateral \textsc{cnot} operation, both target qubits (3 and 4) are measured in the logical basis. If the outcomes of Alice's and Bob's measurement are the same, they keep the source pair, otherwise it is discarded. The fidelity $F'$ of the remaining pair turns out to be larger than the initial fidelity $F$ provided that $1/2 < F < 1$. Another  $\pi$ rotation about the $y$ axis again exchanges $\ket{\Psi^-}$ and $\ket{\Phi^+}$ components. Therefore, iterating the scheme can bring the fidelity arbitrarily close to 1, resulting in a maximally entangled Bell state. Since the state after one purification round is not a Werner state, the last step is necessary as a prerequisite for the twirl in the subsequent round.

The difference in the Deutsch \textit{et al}.~\cite{PhysRevLett.77.2818} protocol (\textsc{dejmps}) is that it works generally on Bell-diagonal states and therefore does not need the twirl to come back to Werner form. A purification round begins with Alice performing the single-qubit gate
\begin{equation}
\label{eq:dejmps}
\ket{0} \mapsto \frac{1}{\sqrt{2}} \big( \ket{0} - i \ket{1} \big), \,
\ket{1} \mapsto \frac{1}{\sqrt{2}} \big( \ket{1} - i \ket{0} \big),
\end{equation}
on both of her qubits and Bob the inverse operation. As in the \textsc{bbpssw} protocol, a bilateral \textsc{cnot} and the measurement of the target pair follow, keeping the source qubits only if the outcomes are equal. If the initial $\ket{\Phi^+}$ component is larger than $1/2$, the states can be purified to a pure $\ket{\Phi^+}$ state, but in a more efficient way than in \cite{PhysRevLett.76.722}.

\textit{Spin Qubits.}
The \textsc{cnot} gate is not directly generated by the typical interaction between spin qubits in gate-defined quantum dots, where the exchange interaction can be described by a Heisenberg Hamiltonian \cite{PhysRevA.57.120,PhysRevB.59.2070}. Two interaction pulses and additional single-qubit operations on both qubits are necessary to construct the \textsc{cnot} gate \cite{PhysRevA.57.120}. 
However, since single-spin rotations are much slower than spin-spin exchange interactions, such an implementation is challenging. 
We therefore construct a purification protocol requiring only a single two-qubit operation each for Alice and Bob, 
which is directly generated from the Heisenberg Hamiltonian describing the exchange interaction
between the spins of two electrons confined to adjacent gate-defined QDs \cite{PhysRevA.57.120,PhysRevB.59.2070},
\begin{equation}
\label{eq:H_Heis}
H_{ij} (t)
=
\frac{1}{4} J(t) \bm{\sigma}_i \cdot \bm{\sigma}_j
=J(t)
\left(
\frac{1}{4}\mathbb{1} - P_{\Psi^-}
\right),
\end{equation}
where $\bm{\sigma}_i$ is the Pauli spin operator of the electron in QD $i$, $P_{\Psi^-} = \ket{\Psi^-}\bra{\Psi^-}$,
and $J(t)$ is the exchange energy between the two electrons that can be tuned by changing appropriate gate voltages, 
and hence can depend on time $t$.
The two possible spin orientations of an electron define the logical states 
$\ket{0} \equiv \ket{\!\uparrow}$ and $\ket{1} \equiv \ket{\! \downarrow}$. 

The time evolution $U$ generated by (\ref{eq:H_Heis}) is given by 
$U_{ij} (\alpha) = \exp [-i \int_0^\tau \text{d}t  \, H_{ij} (t)]
= e^{-i \alpha/4}(\mathbb{1} +(e^{i \alpha} - 1) P_{\Psi^-})$
where we set $\hbar = 1$, assume a pulsed exchange coupling 
of duration $\tau$, and refer to $\alpha = \int_0^\tau \text{d}t \, J(t)$ as the pulse area.
For each value of $\alpha$, a specific two-qubit gate is generated, 
especially the entangling $\sqrt{\textsc{swap}}$ gate for $\alpha = \pi/2$ \cite{PhysRevA.57.120,PhysRevB.59.2070},
$U_{\sqrt{\textsc{swap}}}=e^{i \pi/8} U_{ij} (\pi / 2)$,
which together with arbitrary single-qubit operations forms a universal set of quantum gates \cite{PhysRevA.57.120,PhysRevA.52.3457}, or the \textsc{swap} operation for $\alpha = \pi$,
$U_\textsc{swap}=( U_{\sqrt{\textsc{swap}}} )^2=e^{i \pi/4} U_{ij} (\pi)$,
whose action is to interchange the states of the qubits \cite{nielsen00}. The \textsc{cnot} can be obtained by the sequence \cite{PhysRevA.57.120}
$U_\textsc{cnot}=
e^{- i \pi/2} e^{- i \pi \sigma_2^y/4} e^{i \pi \sigma_1^z/4} e^{- i \pi \sigma_2^z/4} U_{\!\!\sqrt{\textsc{swap}}} e^{i \pi \sigma_1^z/2} U_{\!\!\sqrt{\textsc{swap}}} e^{i \pi \sigma_2^y/4}$, requiring two $\sqrt{\textsc{swap}}$ and several single-qubit gates.

\textit{Purification with the Exchange Interaction.}
We now introduce a purification protocol, that only makes use of a single activation of the Heisenberg interaction (\ref{eq:H_Heis}). 
The protocol has the same structure as existing recurrence protocols \cite{PhysRevLett.76.722}, the crucial difference being that the bilateral operation is \textit{asymmetric}, meaning one has to apply \textit{different} two-qubit gates.
The two parties Alice and Bob each start with a copy of the state $\rho_F$,
with $\rho=\rho_F \otimes \rho_F$,
which can be generated as shown above.
Then, Alice and Bob each activate the exchange interaction between their two qubits with specific pulse areas $\alpha$ and $\beta$ respectively (Fig.~\ref{fig:circuit}), i.e.~they apply the unitary transformation
\begin{equation}
U(\alpha,\beta)
=
U_{13} (\alpha) \otimes U_{24} (\beta),
\end{equation}
where Alice holds qubits 1 and 3, Bob qubits 2 and 4. This transforms the initial state $\rho$ into
$U(\alpha,\beta) \rho U(\alpha,\beta)^\dagger$.
After this unitary transformation, the two parties continue as in the \textsc{bbpssw} protocol. The target qubits 3 and 4 
are measured in the $z$ basis and the results are communicated via a classical channel. 
If the spins point in the same direction, Alice and Bob keep qubits 1 and 2 (source pair), otherwise the state is discarded.

We find the fidelity $F' = \bra{\Phi^+} \rho' \ket{\Phi^+}$ of the postselected source state $\rho'$ after the described procedure to be
\begin{widetext}
\begin{equation}
\label{eq:fidelity}
F' (F, \alpha, \beta)
=
\frac{(4F-1)(4F+5) \cos \alpha \cos \beta - (4F-1)(8F+1) \sin \alpha \sin \beta + 8F(4F+1) + 5}
{6(4F-1) \cos \alpha \cos \beta - 2(4F-1)^2 \sin \alpha \sin \beta + 6(4F+5)} ,
\end{equation}
\end{widetext}
which is the main result of this paper.
\begin{figure}[b]
\includegraphics[width=\columnwidth]{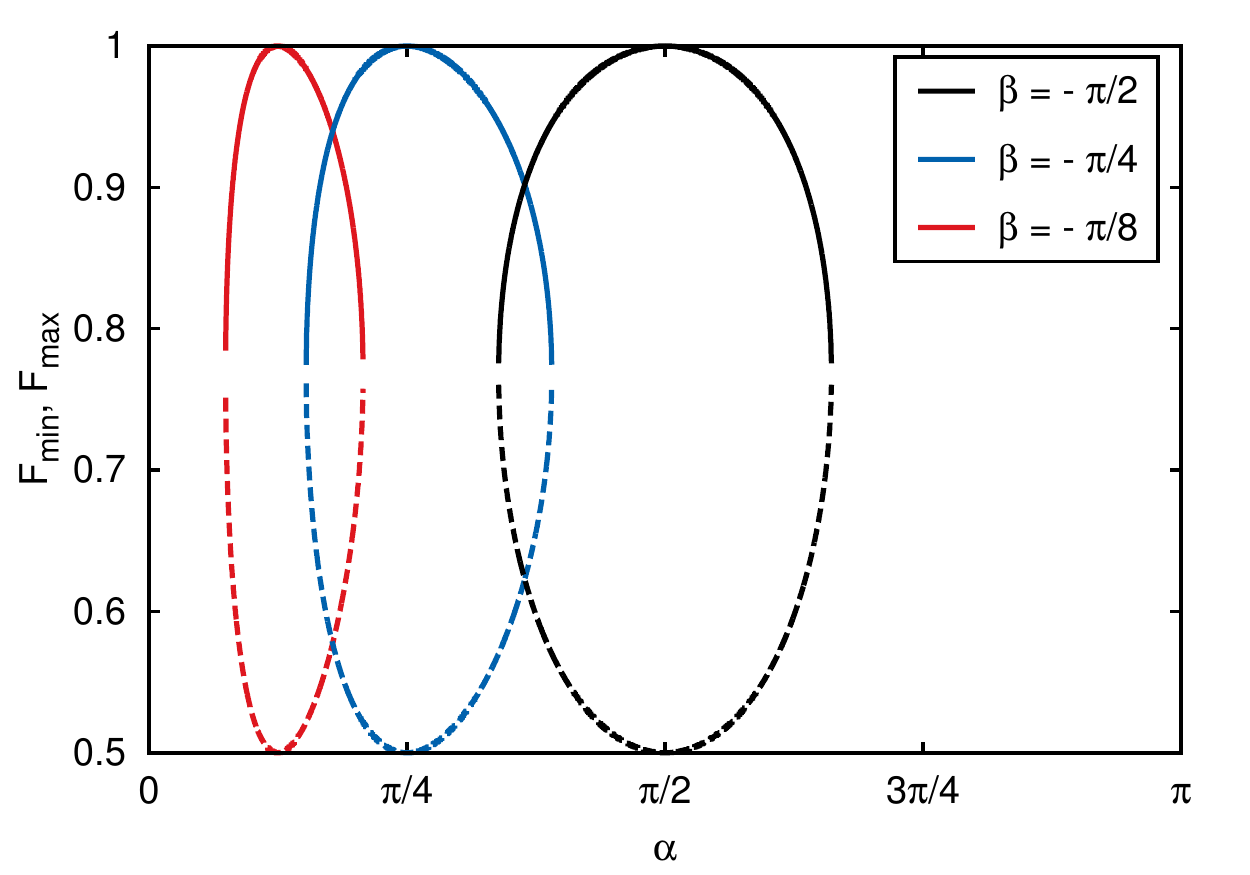}
\caption{\label{fig:fixed_points} The minimum required initial fidelity $F_\text{min}$ (dashed line) for the entanglement purification to work and the maximally attainable fidelity $F_\text{max}$ (solid line) when iteratively applying the purification protocol generating the map (\ref{eq:fidelity}) as a function of the pulse area $\alpha$, for different pulse areas $\beta$.}
\end{figure}

To show the feasibility of the protocol, the three fixed points of the map (\ref{eq:fidelity}) can be found analytically. Except for the case $\alpha = n\pi$, $\beta = m\pi$, and $\alpha - \beta = 2 \pi k$ ($n,m,k$ integers), where $F' \equiv F$, a constant fixed point is given by $F_\text{c} = 1/4$. The values of the two remaining (possibly complex) fixed points $F_\text{min}$ and $F_\text{max}$ depend on $\alpha$ and $\beta$, as illustrated in Fig.~\ref{fig:fixed_points} within the physically relevant regime $1/2\leq F_\text{min} \leq F_\text{max} \leq 1$. While $F_\text{max}$ and $F_\text{c}$ are attractive, $F_\text{min}$ is repulsive. Thus, if the qubit pairs have an initial fidelity $F>F_\text{min}\ge 1/2$, iterative application of the described scheme can purify them up to a fidelity $F_\text{max}$.

\begin{figure}
\includegraphics[width=\columnwidth]{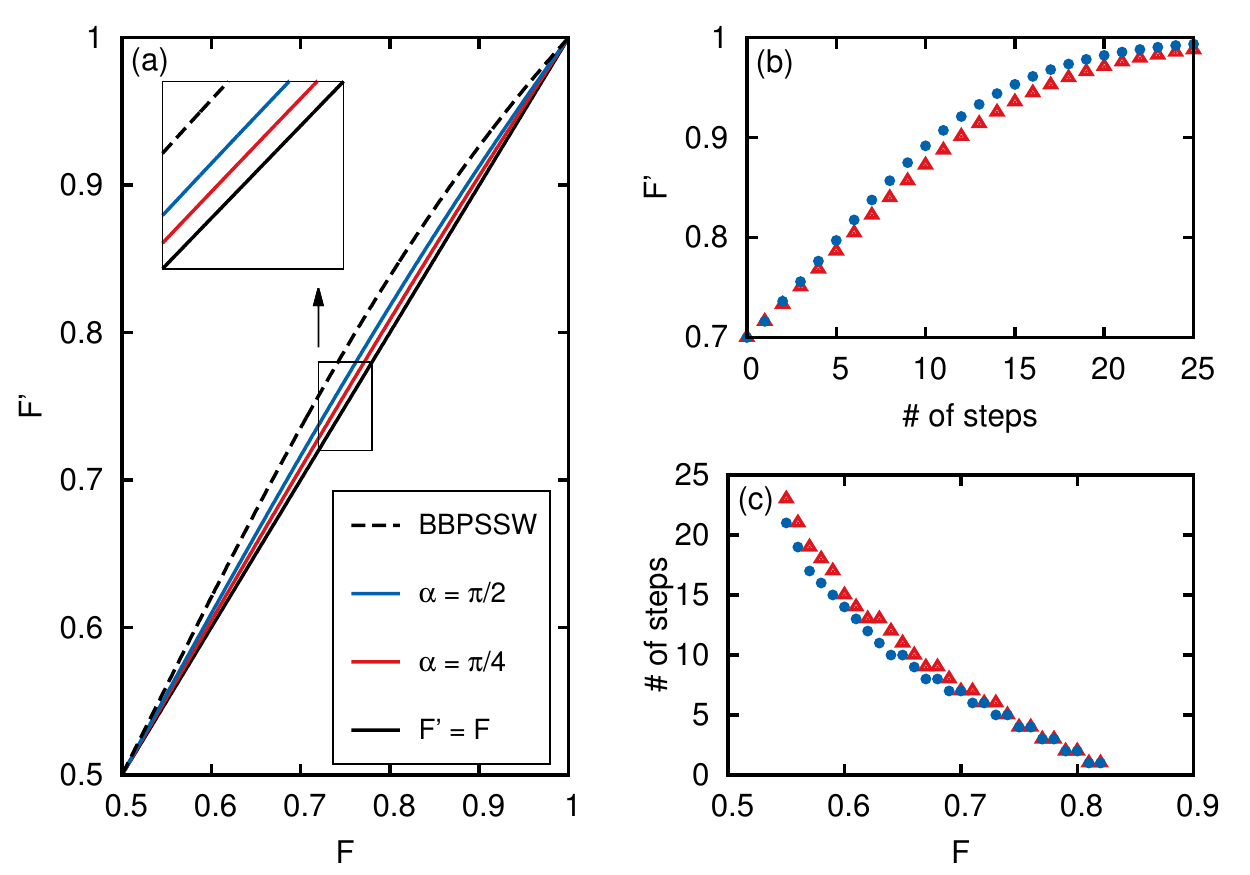}
\caption{\label{fig:fiidelity_special_case} (a) The fidelity $F'$ as a function of the initial fidelity $F$, shown for different pulse areas $\alpha$ and compared to the \textsc{bbpssw} protocol. (b) Stepwise fidelity increase for an initial fidelity $F=0.7$ and (c) number of steps needed to achieve a fidelity $F' > 0.99$ starting from qubit pairs with a fidelity $F$; shown for the described protocol (triangles) in the optimal case (\ref{eq:fid_optimal}) and in the more efficient way in analogy to the \textsc{dejmps} protocol (dots).}
\end{figure}

\textit{Optimal Purification.}
The minimal and maximal values of the fixed points $F_\text{min} = 1/2$ and $F_\text{max} = 1$ are obtained in the case $\beta = - \alpha$. This implies that in principle maximally entangled states can be created using our protocol if Alice and Bob perform mutually inverse operations. This could be achieved either using ferro- and antiferromagnetic exchange $J_A>0$, $J_B<0$, or vice versa, or in the case of equal coupling types, using the fact that the propagator $U_{ij}$ is $2 \pi$ periodic (omitting overall phases), with, e.g., $0<\alpha<\pi$ and $\beta = 2\pi -\alpha>0$.
The fidelity is then given by
\begin{eqnarray}
\label{eq:fid_special}
&&F'(F, \alpha, - \alpha) = \nonumber \\
&&\frac{1}{2} + \frac{3 - 12 F^2}{(F - 1)(4F - 1) \cos(2 \alpha) - F(4F + 7) - 7},
\end{eqnarray}
and has its maximum in the optimal case $\alpha = \pi/2$,
\begin{equation}
\label{eq:fid_optimal}
F'
\left(
F, \frac{\pi}{2} + 2\pi n, - \frac{\pi}{2} + 2\pi m
\right)
=
\frac{16 F^2 + F + 1}{8 F^2 + 2 F + 8}.
\end{equation}
The result Eq.~(\ref{eq:fid_special}) is plotted in Fig.~\ref{fig:fiidelity_special_case}(a) for different values of $\alpha$. The optimal value is therefore achieved if Alice applies a $\sqrt{\textsc{swap}}$ gate to her qubits and Bob performs the inverse $\sqrt{\textsc{swap}}$ gate, $\sqrt{\textsc{swap}}^{-1}$. 
The square of  $\sqrt{\textsc{swap}}^{-1}$ is also the \textsc{swap} operation and it can be understood as another root of \textsc{swap}.

The described protocol can be slightly improved in terms of resources needed to achieve a specific fidelity. In analogy to the \textsc{dejmps} protocol, Alice can begin each purification step with the local operations (\ref{eq:dejmps}) (and Bob with the inverse) and the bilateral \textsc{cnot} is replaced by the asymmetric operation described above, to increase the gain in fidelity (Fig.~\ref{fig:fiidelity_special_case}(b)) and therefore reduce the number of purification steps (Fig.~\ref{fig:fiidelity_special_case}(c)).
Furthermore, one can see from Fig.~\ref{fig:fixed_points} that the robustness of our scheme is largest in the case $\alpha = - \beta =\pi / 2$, since deviations from perfectly applied operations have the least effect on the values of $F_\text{min}$ and $F_\text{max}$. If Alice applies a rectangular pulse with an exchange interaction of $J= 1 \, \mu \text{eV}$ \cite{Petta30092005}, the deviation $\Delta \tau$ from the optimal case must not exceed 100 ps to still achieve a maximum fidelity of 0.99, assuming Bob generates a perfect pulse. Such accuracies can be obtained experimentally  \cite{Petta30092005}.

\textit{Purification with the XY Interaction.}
We briefly discuss our approach for entanglement purification and applying an asymmetric bilateral operation for the case of anisotropic XY-type qubit interactions, $H_{\textsc{xy}} (t)
=
\frac{1}{4} J(t)
\left(
\sigma_1^x \sigma_2^x + \sigma_1^y \sigma_2^y
\right)$.
This kind of interaction appears, e.g., in all-optical cavity-coupled QD electron spins \cite{PhysRevLett.83.4204} or superconducting qubits \cite{RevModPhys.73.357}.
The Hamiltonian $H_{\textsc{xy}} (t)$ generates the i\textsc{swap} gate,
$\ket{\! \uparrow \uparrow} \mapsto \ket{\! \uparrow \uparrow}$,
$\ket{\! \uparrow \downarrow} \mapsto i \ket{\! \downarrow \uparrow}$,
$\ket{\! \downarrow \uparrow} \mapsto i \ket{\! \uparrow \downarrow}$,
$\ket{\! \downarrow \downarrow} \mapsto \ket{\! \downarrow \downarrow}$,
for a pulse area $\int_0^\tau \text{d}t' J(t') = - \pi$.
Following the above scheme, applying interactions with different pulse areas  $\alpha$ and  $\beta$ on Alice's and Bob's
qubits, the fidelity $F'$ after the protocol is
\begin{widetext}
\begin{equation}
F'(F, \alpha, \beta)
=
\frac{(12F-3) \cos \alpha \cos \beta - (4F-1)^2 \sin \alpha \sin \beta + 4(8F^2 + 2F -1)\cos\frac{\alpha + \beta}{2} + 4F(4F+1) + 7}
{6(4F-1) \cos \alpha \cos \beta - 2(4F-1)^2\sin \alpha \sin \beta + 6(4F+5)}.
\end{equation}
\end{widetext}
In the case $\alpha = - \beta$, the result coincides with (\ref{eq:fid_special}) and therefore is maximal for $\alpha = \pi / 2$. 
The optimal bilateral two-qubit gates in the case of the XY Hamiltonian are square roots of i\textsc{swap}.

\textit{Conclusions.}
We have presented an entanglement purification scheme, in which the bilateral two-qubit operation is directly generated from the one-time activation of a Heisenberg-type spin-spin interaction. In general we could show that an asymmetric unitary evolution of Alice's and Bob's qubits, respectively, can lead to an increased fidelity of one of the shared qubit pairs with respect to $\ket{\Phi^+}$ if the initial fidelity was larger than a given minimal value. In the special case where Alice and Bob apply inverse operations, the maximally obtainable fidelity by iterative application of our protocol 
is $F=1$, i.e., in principle maximally entangled states can be generated. We found that the optimal case is when the two communicating parties apply the $\sqrt{\textsc{swap}}$ and the $\sqrt{\textsc{swap}}^{-1}$ gates locally on their qubits.

Since the coupling of electron spins in gate-controlled QDs is well described by an exchange interaction of Heisenberg type, the protocol is particularly  suitable for spin qubits. In terms of operation times, the presented protocol is much faster than protocols based on \textsc{cnot} applied to spin qubits. The reason is that single-qubit gates needed in constructing the \textsc{cnot}  need operation times on the order of 100 ns \cite{Nowack30112007} whereas the $\sqrt{\textsc{swap}}$ can be generated two orders of magnitude faster in about 0.2 ns \cite{Petta30092005}. Therefore, besides the smaller error-proneness due to the smaller number of gate operations, a much faster iteration of the protocol is possible. In addition, we showed that our purification method of applying an asymmetric bilateral operation works as well for qubits coupled via a XY-type interaction and is therefore suitable for cavity-coupled spin qubits and superconducting qubits.

\textit{Acknowledgements.}
We acknowledge funding from the DFG within SFB 767 and from the BMBF under the program QuaHL-Rep.

\bibliography{bibliography-all}

\begin{thebibliography}{31}%
\makeatletter
\providecommand \@ifxundefined [1]{%
 \@ifx{#1\undefined}
}%
\providecommand \@ifnum [1]{%
 \ifnum #1\expandafter \@firstoftwo
 \else \expandafter \@secondoftwo
 \fi
}%
\providecommand \@ifx [1]{%
 \ifx #1\expandafter \@firstoftwo
 \else \expandafter \@secondoftwo
 \fi
}%
\providecommand \natexlab [1]{#1}%
\providecommand \enquote  [1]{``#1''}%
\providecommand \bibnamefont  [1]{#1}%
\providecommand \bibfnamefont [1]{#1}%
\providecommand \citenamefont [1]{#1}%
\providecommand \href@noop [0]{\@secondoftwo}%
\providecommand \href [0]{\begingroup \@sanitize@url \@href}%
\providecommand \@href[1]{\@@startlink{#1}\@@href}%
\providecommand \@@href[1]{\endgroup#1\@@endlink}%
\providecommand \@sanitize@url [0]{\catcode `\\12\catcode `\$12\catcode
  `\&12\catcode `\#12\catcode `\^12\catcode `\_12\catcode `\%12\relax}%
\providecommand \@@startlink[1]{}%
\providecommand \@@endlink[0]{}%
\providecommand \url  [0]{\begingroup\@sanitize@url \@url }%
\providecommand \@url [1]{\endgroup\@href {#1}{\urlprefix }}%
\providecommand \urlprefix  [0]{URL }%
\providecommand \Eprint [0]{\href }%
\providecommand \doibase [0]{http://dx.doi.org/}%
\providecommand \selectlanguage [0]{\@gobble}%
\providecommand \bibinfo  [0]{\@secondoftwo}%
\providecommand \bibfield  [0]{\@secondoftwo}%
\providecommand \translation [1]{[#1]}%
\providecommand \BibitemOpen [0]{}%
\providecommand \bibitemStop [0]{}%
\providecommand \bibitemNoStop [0]{.\EOS\space}%
\providecommand \EOS [0]{\spacefactor3000\relax}%
\providecommand \BibitemShut  [1]{\csname bibitem#1\endcsname}%
\let\auto@bib@innerbib\@empty
\bibitem [{\citenamefont {Bouwmeester}\ \emph {et~al.}(2001)\citenamefont
  {Bouwmeester}, \citenamefont {Ekert},\ and\ \citenamefont
  {Zeilinger}}]{bouwmeester2000physics}%
  \BibitemOpen
  \bibfield  {author} {\bibinfo {author} {\bibfnamefont {D.}~\bibnamefont
  {Bouwmeester}}, \bibinfo {author} {\bibfnamefont {A.~K.}\ \bibnamefont
  {Ekert}}, \ and\ \bibinfo {author} {\bibfnamefont {A.}~\bibnamefont
  {Zeilinger}},\ }\href@noop {} {\emph {\bibinfo {title} {The Physics of
  Quantum Information: Quantum Cryptography, Quantum Teleportation, Quantum
  Computation}}}\ (\bibinfo  {publisher} {Springer New York},\ \bibinfo {year}
  {2001})\BibitemShut {NoStop}%
\bibitem [{\citenamefont {Bennett}\ \emph {et~al.}(1993)\citenamefont
  {Bennett}, \citenamefont {Brassard}, \citenamefont {Cr\'epeau}, \citenamefont
  {Jozsa}, \citenamefont {Peres},\ and\ \citenamefont
  {Wootters}}]{PhysRevLett.70.1895}%
  \BibitemOpen
  \bibfield  {author} {\bibinfo {author} {\bibfnamefont {C.~H.}\ \bibnamefont
  {Bennett}}, \bibinfo {author} {\bibfnamefont {G.}~\bibnamefont {Brassard}},
  \bibinfo {author} {\bibfnamefont {C.}~\bibnamefont {Cr\'epeau}}, \bibinfo
  {author} {\bibfnamefont {R.}~\bibnamefont {Jozsa}}, \bibinfo {author}
  {\bibfnamefont {A.}~\bibnamefont {Peres}}, \ and\ \bibinfo {author}
  {\bibfnamefont {W.~K.}\ \bibnamefont {Wootters}},\ }\href {\doibase
  10.1103/PhysRevLett.70.1895} {\bibfield  {journal} {\bibinfo  {journal}
  {Phys. Rev. Lett.}\ }\textbf {\bibinfo {volume} {70}},\ \bibinfo {pages}
  {1895} (\bibinfo {year} {1993})}\BibitemShut {NoStop}%
\bibitem [{\citenamefont {Bennett}\ and\ \citenamefont
  {Wiesner}(1992)}]{PhysRevLett.69.2881}%
  \BibitemOpen
  \bibfield  {author} {\bibinfo {author} {\bibfnamefont {C.~H.}\ \bibnamefont
  {Bennett}}\ and\ \bibinfo {author} {\bibfnamefont {S.~J.}\ \bibnamefont
  {Wiesner}},\ }\href {\doibase 10.1103/PhysRevLett.69.2881} {\bibfield
  {journal} {\bibinfo  {journal} {Phys. Rev. Lett.}\ }\textbf {\bibinfo
  {volume} {69}},\ \bibinfo {pages} {2881} (\bibinfo {year}
  {1992})}\BibitemShut {NoStop}%
\bibitem [{\citenamefont {Bennett}\ and\ \citenamefont
  {Brassard}(1984)}]{BB84}%
  \BibitemOpen
  \bibfield  {author} {\bibinfo {author} {\bibfnamefont {C.~H.}\ \bibnamefont
  {Bennett}}\ and\ \bibinfo {author} {\bibfnamefont {G.}~\bibnamefont
  {Brassard}},\ }in\ \href@noop {} {\emph {\bibinfo {booktitle} {Proceedings of
  the IEEE International Conference on Computers, Systems, and Signal
  Processing, Bangalore}}}\ (\bibinfo {year} {1984})\ pp.\ \bibinfo {pages}
  {175--179}\BibitemShut {NoStop}%
\bibitem [{\citenamefont {Ekert}(1991)}]{PhysRevLett.67.661}%
  \BibitemOpen
  \bibfield  {author} {\bibinfo {author} {\bibfnamefont {A.~K.}\ \bibnamefont
  {Ekert}},\ }\href {\doibase 10.1103/PhysRevLett.67.661} {\bibfield  {journal}
  {\bibinfo  {journal} {Phys. Rev. Lett.}\ }\textbf {\bibinfo {volume} {67}},\
  \bibinfo {pages} {661} (\bibinfo {year} {1991})}\BibitemShut {NoStop}%
\bibitem [{\citenamefont {Bennett}\ \emph
  {et~al.}(1996{\natexlab{a}})\citenamefont {Bennett}, \citenamefont
  {Brassard}, \citenamefont {Popescu}, \citenamefont {Schumacher},
  \citenamefont {Smolin},\ and\ \citenamefont {Wootters}}]{PhysRevLett.76.722}%
  \BibitemOpen
  \bibfield  {author} {\bibinfo {author} {\bibfnamefont {C.~H.}\ \bibnamefont
  {Bennett}}, \bibinfo {author} {\bibfnamefont {G.}~\bibnamefont {Brassard}},
  \bibinfo {author} {\bibfnamefont {S.}~\bibnamefont {Popescu}}, \bibinfo
  {author} {\bibfnamefont {B.}~\bibnamefont {Schumacher}}, \bibinfo {author}
  {\bibfnamefont {J.~A.}\ \bibnamefont {Smolin}}, \ and\ \bibinfo {author}
  {\bibfnamefont {W.~K.}\ \bibnamefont {Wootters}},\ }\href {\doibase
  10.1103/PhysRevLett.76.722} {\bibfield  {journal} {\bibinfo  {journal} {Phys.
  Rev. Lett.}\ }\textbf {\bibinfo {volume} {76}},\ \bibinfo {pages} {722}
  (\bibinfo {year} {1996}{\natexlab{a}})}\BibitemShut {NoStop}%
\bibitem [{\citenamefont {Deutsch}\ \emph {et~al.}(1996)\citenamefont
  {Deutsch}, \citenamefont {Ekert}, \citenamefont {Jozsa}, \citenamefont
  {Macchiavello}, \citenamefont {Popescu},\ and\ \citenamefont
  {Sanpera}}]{PhysRevLett.77.2818}%
  \BibitemOpen
  \bibfield  {author} {\bibinfo {author} {\bibfnamefont {D.}~\bibnamefont
  {Deutsch}}, \bibinfo {author} {\bibfnamefont {A.}~\bibnamefont {Ekert}},
  \bibinfo {author} {\bibfnamefont {R.}~\bibnamefont {Jozsa}}, \bibinfo
  {author} {\bibfnamefont {C.}~\bibnamefont {Macchiavello}}, \bibinfo {author}
  {\bibfnamefont {S.}~\bibnamefont {Popescu}}, \ and\ \bibinfo {author}
  {\bibfnamefont {A.}~\bibnamefont {Sanpera}},\ }\href {\doibase
  10.1103/PhysRevLett.77.2818} {\bibfield  {journal} {\bibinfo  {journal}
  {Phys. Rev. Lett.}\ }\textbf {\bibinfo {volume} {77}},\ \bibinfo {pages}
  {2818} (\bibinfo {year} {1996})}\BibitemShut {NoStop}%
\bibitem [{\citenamefont {D{\"u}r}\ and\ \citenamefont
  {Briegel}(2007)}]{0034-4885-70-8-R03}%
  \BibitemOpen
  \bibfield  {author} {\bibinfo {author} {\bibfnamefont {W.}~\bibnamefont
  {D{\"u}r}}\ and\ \bibinfo {author} {\bibfnamefont {H.~J.}\ \bibnamefont
  {Briegel}},\ }\href {http://stacks.iop.org/0034-4885/70/i=8/a=R03} {\bibfield
   {journal} {\bibinfo  {journal} {Rep. Prog. Phys.}\ }\textbf {\bibinfo
  {volume} {70}},\ \bibinfo {pages} {1381} (\bibinfo {year}
  {2007})}\BibitemShut {NoStop}%
\bibitem [{\citenamefont {Briegel}\ \emph {et~al.}(1998)\citenamefont
  {Briegel}, \citenamefont {D\"ur}, \citenamefont {Cirac},\ and\ \citenamefont
  {Zoller}}]{PhysRevLett.81.5932}%
  \BibitemOpen
  \bibfield  {author} {\bibinfo {author} {\bibfnamefont {H.-J.}\ \bibnamefont
  {Briegel}}, \bibinfo {author} {\bibfnamefont {W.}~\bibnamefont {D\"ur}},
  \bibinfo {author} {\bibfnamefont {J.~I.}\ \bibnamefont {Cirac}}, \ and\
  \bibinfo {author} {\bibfnamefont {P.}~\bibnamefont {Zoller}},\ }\href
  {\doibase 10.1103/PhysRevLett.81.5932} {\bibfield  {journal} {\bibinfo
  {journal} {Phys. Rev. Lett.}\ }\textbf {\bibinfo {volume} {81}},\ \bibinfo
  {pages} {5932} (\bibinfo {year} {1998})}\BibitemShut {NoStop}%
\bibitem [{\citenamefont {D\"ur}\ \emph {et~al.}(1999)\citenamefont {D\"ur},
  \citenamefont {Briegel}, \citenamefont {Cirac},\ and\ \citenamefont
  {Zoller}}]{PhysRevA.59.169}%
  \BibitemOpen
  \bibfield  {author} {\bibinfo {author} {\bibfnamefont {W.}~\bibnamefont
  {D\"ur}}, \bibinfo {author} {\bibfnamefont {H.-J.}\ \bibnamefont {Briegel}},
  \bibinfo {author} {\bibfnamefont {J.~I.}\ \bibnamefont {Cirac}}, \ and\
  \bibinfo {author} {\bibfnamefont {P.}~\bibnamefont {Zoller}},\ }\href
  {\doibase 10.1103/PhysRevA.59.169} {\bibfield  {journal} {\bibinfo  {journal}
  {Phys. Rev. A}\ }\textbf {\bibinfo {volume} {59}},\ \bibinfo {pages} {169}
  (\bibinfo {year} {1999})}\BibitemShut {NoStop}%
\bibitem [{\citenamefont {Kimble}(2008)}]{Kimble:2008}%
  \BibitemOpen
  \bibfield  {author} {\bibinfo {author} {\bibfnamefont {H.~J.}\ \bibnamefont
  {Kimble}},\ }\href {http://dx.doi.org/10.1038/nature07127} {\bibfield
  {journal} {\bibinfo  {journal} {Nature}\ }\textbf {\bibinfo {volume} {453}},\
  \bibinfo {pages} {1023} (\bibinfo {year} {2008})}\BibitemShut {NoStop}%
\bibitem [{\citenamefont {Simon}\ \emph {et~al.}(2010)\citenamefont {Simon},
  \citenamefont {Afzelius}, \citenamefont {Appel}, \citenamefont {Boyer de~la
  Giroday}, \citenamefont {Dewhurst}, \citenamefont {Gisin}, \citenamefont
  {Hu}, \citenamefont {Jelezko}, \citenamefont {Kr\"oll}, \citenamefont
  {M\"uller}, \citenamefont {Nunn}, \citenamefont {Polzik}, \citenamefont
  {Rarity}, \citenamefont {De~Riedmatten}, \citenamefont {Rosenfeld},
  \citenamefont {Shields}, \citenamefont {Sk\"old}, \citenamefont {Stevenson},
  \citenamefont {Thew}, \citenamefont {Walmsley}, \citenamefont {Weber},
  \citenamefont {Weinfurter}, \citenamefont {Wrachtrup},\ and\ \citenamefont
  {Young}}]{Simon:2010}%
  \BibitemOpen
  \bibfield  {author} {\bibinfo {author} {\bibfnamefont {C.}~\bibnamefont
  {Simon}}, \bibinfo {author} {\bibfnamefont {M.}~\bibnamefont {Afzelius}},
  \bibinfo {author} {\bibfnamefont {J.}~\bibnamefont {Appel}}, \bibinfo
  {author} {\bibfnamefont {A.}~\bibnamefont {Boyer de~la Giroday}}, \bibinfo
  {author} {\bibfnamefont {S.~J.}\ \bibnamefont {Dewhurst}}, \bibinfo {author}
  {\bibfnamefont {N.}~\bibnamefont {Gisin}}, \bibinfo {author} {\bibfnamefont
  {C.~Y.}\ \bibnamefont {Hu}}, \bibinfo {author} {\bibfnamefont
  {F.}~\bibnamefont {Jelezko}}, \bibinfo {author} {\bibfnamefont
  {S.}~\bibnamefont {Kr\"oll}}, \bibinfo {author} {\bibfnamefont {J.~H.}\
  \bibnamefont {M\"uller}}, \bibinfo {author} {\bibfnamefont {J.}~\bibnamefont
  {Nunn}}, \bibinfo {author} {\bibfnamefont {E.~S.}\ \bibnamefont {Polzik}},
  \bibinfo {author} {\bibfnamefont {J.~G.}\ \bibnamefont {Rarity}}, \bibinfo
  {author} {\bibfnamefont {H.}~\bibnamefont {De~Riedmatten}}, \bibinfo {author}
  {\bibfnamefont {W.}~\bibnamefont {Rosenfeld}}, \bibinfo {author}
  {\bibfnamefont {A.~J.}\ \bibnamefont {Shields}}, \bibinfo {author}
  {\bibfnamefont {N.}~\bibnamefont {Sk\"old}}, \bibinfo {author} {\bibfnamefont
  {R.~M.}\ \bibnamefont {Stevenson}}, \bibinfo {author} {\bibfnamefont
  {R.}~\bibnamefont {Thew}}, \bibinfo {author} {\bibfnamefont {I.~A.}\
  \bibnamefont {Walmsley}}, \bibinfo {author} {\bibfnamefont {M.~C.}\
  \bibnamefont {Weber}}, \bibinfo {author} {\bibfnamefont {H.}~\bibnamefont
  {Weinfurter}}, \bibinfo {author} {\bibfnamefont {J.}~\bibnamefont
  {Wrachtrup}}, \ and\ \bibinfo {author} {\bibfnamefont {R.~J.}\ \bibnamefont
  {Young}},\ }\href {\doibase 10.1140/epjd/e2010-00103-y} {\bibfield  {journal}
  {\bibinfo  {journal} {Eur. Phys. J. D}\ }\textbf {\bibinfo {volume} {58}},\
  \bibinfo {pages} {1} (\bibinfo {year} {2010})}\BibitemShut {NoStop}%
\bibitem [{\citenamefont {\ifmmode~\dot{Z}\else \.{Z}\fi{}ukowski}\ \emph
  {et~al.}(1993)\citenamefont {\ifmmode~\dot{Z}\else \.{Z}\fi{}ukowski},
  \citenamefont {Zeilinger}, \citenamefont {Horne},\ and\ \citenamefont
  {Ekert}}]{PhysRevLett.71.4287}%
  \BibitemOpen
  \bibfield  {author} {\bibinfo {author} {\bibfnamefont {M.}~\bibnamefont
  {\ifmmode~\dot{Z}\else \.{Z}\fi{}ukowski}}, \bibinfo {author} {\bibfnamefont
  {A.}~\bibnamefont {Zeilinger}}, \bibinfo {author} {\bibfnamefont {M.~A.}\
  \bibnamefont {Horne}}, \ and\ \bibinfo {author} {\bibfnamefont {A.~K.}\
  \bibnamefont {Ekert}},\ }\href {\doibase 10.1103/PhysRevLett.71.4287}
  {\bibfield  {journal} {\bibinfo  {journal} {Phys. Rev. Lett.}\ }\textbf
  {\bibinfo {volume} {71}},\ \bibinfo {pages} {4287} (\bibinfo {year}
  {1993})}\BibitemShut {NoStop}%
\bibitem [{\citenamefont {Loss}\ and\ \citenamefont
  {DiVincenzo}(1998)}]{PhysRevA.57.120}%
  \BibitemOpen
  \bibfield  {author} {\bibinfo {author} {\bibfnamefont {D.}~\bibnamefont
  {Loss}}\ and\ \bibinfo {author} {\bibfnamefont {D.~P.}\ \bibnamefont
  {DiVincenzo}},\ }\href {\doibase 10.1103/PhysRevA.57.120} {\bibfield
  {journal} {\bibinfo  {journal} {Phys. Rev. A}\ }\textbf {\bibinfo {volume}
  {57}},\ \bibinfo {pages} {120} (\bibinfo {year} {1998})}\BibitemShut
  {NoStop}%
\bibitem [{\citenamefont {Koppens}\ \emph {et~al.}(2008)\citenamefont
  {Koppens}, \citenamefont {Nowack},\ and\ \citenamefont
  {Vandersypen}}]{PhysRevLett.100.236802}%
  \BibitemOpen
  \bibfield  {author} {\bibinfo {author} {\bibfnamefont {F.~H.~L.}\
  \bibnamefont {Koppens}}, \bibinfo {author} {\bibfnamefont {K.~C.}\
  \bibnamefont {Nowack}}, \ and\ \bibinfo {author} {\bibfnamefont {L.~M.~K.}\
  \bibnamefont {Vandersypen}},\ }\href {\doibase
  10.1103/PhysRevLett.100.236802} {\bibfield  {journal} {\bibinfo  {journal}
  {Phys. Rev. Lett.}\ }\textbf {\bibinfo {volume} {100}},\ \bibinfo {pages}
  {236802} (\bibinfo {year} {2008})}\BibitemShut {NoStop}%
\bibitem [{\citenamefont {Hanson}\ \emph {et~al.}(2007)\citenamefont {Hanson},
  \citenamefont {Kouwenhoven}, \citenamefont {Petta}, \citenamefont {Tarucha},\
  and\ \citenamefont {Vandersypen}}]{RevModPhys.79.1217}%
  \BibitemOpen
  \bibfield  {author} {\bibinfo {author} {\bibfnamefont {R.}~\bibnamefont
  {Hanson}}, \bibinfo {author} {\bibfnamefont {L.~P.}\ \bibnamefont
  {Kouwenhoven}}, \bibinfo {author} {\bibfnamefont {J.~R.}\ \bibnamefont
  {Petta}}, \bibinfo {author} {\bibfnamefont {S.}~\bibnamefont {Tarucha}}, \
  and\ \bibinfo {author} {\bibfnamefont {L.~M.~K.}\ \bibnamefont
  {Vandersypen}},\ }\href {\doibase 10.1103/RevModPhys.79.1217} {\bibfield
  {journal} {\bibinfo  {journal} {Rev. Mod. Phys.}\ }\textbf {\bibinfo {volume}
  {79}},\ \bibinfo {pages} {1217} (\bibinfo {year} {2007})}\BibitemShut
  {NoStop}%
\bibitem [{\citenamefont {Berezovsky}\ \emph {et~al.}(2008)\citenamefont
  {Berezovsky}, \citenamefont {Mikkelsen}, \citenamefont {Stoltz},
  \citenamefont {Coldren},\ and\ \citenamefont
  {Awschalom}}]{Berezovsky18042008}%
  \BibitemOpen
  \bibfield  {author} {\bibinfo {author} {\bibfnamefont {J.}~\bibnamefont
  {Berezovsky}}, \bibinfo {author} {\bibfnamefont {M.~H.}\ \bibnamefont
  {Mikkelsen}}, \bibinfo {author} {\bibfnamefont {N.~G.}\ \bibnamefont
  {Stoltz}}, \bibinfo {author} {\bibfnamefont {L.~A.}\ \bibnamefont {Coldren}},
  \ and\ \bibinfo {author} {\bibfnamefont {D.~D.}\ \bibnamefont {Awschalom}},\
  }\href {\doibase 10.1126/science.1154798} {\bibfield  {journal} {\bibinfo
  {journal} {Science}\ }\textbf {\bibinfo {volume} {320}},\ \bibinfo {pages}
  {349} (\bibinfo {year} {2008})}\BibitemShut {NoStop}%
\bibitem [{\citenamefont {Makhlin}\ \emph {et~al.}(2001)\citenamefont
  {Makhlin}, \citenamefont {Sch\"on},\ and\ \citenamefont
  {Shnirman}}]{RevModPhys.73.357}%
  \BibitemOpen
  \bibfield  {author} {\bibinfo {author} {\bibfnamefont {Y.}~\bibnamefont
  {Makhlin}}, \bibinfo {author} {\bibfnamefont {G.}~\bibnamefont {Sch\"on}}, \
  and\ \bibinfo {author} {\bibfnamefont {A.}~\bibnamefont {Shnirman}},\ }\href
  {\doibase 10.1103/RevModPhys.73.357} {\bibfield  {journal} {\bibinfo
  {journal} {Rev. Mod. Phys.}\ }\textbf {\bibinfo {volume} {73}},\ \bibinfo
  {pages} {357} (\bibinfo {year} {2001})}\BibitemShut {NoStop}%
\bibitem [{\citenamefont {Imamo\ifmmode~\breve{g}\else \u{g}\fi{}lu}\ \emph
  {et~al.}(1999)\citenamefont {Imamo\ifmmode~\breve{g}\else \u{g}\fi{}lu},
  \citenamefont {Awschalom}, \citenamefont {Burkard}, \citenamefont
  {DiVincenzo}, \citenamefont {Loss}, \citenamefont {Sherwin},\ and\
  \citenamefont {Small}}]{PhysRevLett.83.4204}%
  \BibitemOpen
  \bibfield  {author} {\bibinfo {author} {\bibfnamefont {A.}~\bibnamefont
  {Imamo\ifmmode~\breve{g}\else \u{g}\fi{}lu}}, \bibinfo {author}
  {\bibfnamefont {D.~D.}\ \bibnamefont {Awschalom}}, \bibinfo {author}
  {\bibfnamefont {G.}~\bibnamefont {Burkard}}, \bibinfo {author} {\bibfnamefont
  {D.~P.}\ \bibnamefont {DiVincenzo}}, \bibinfo {author} {\bibfnamefont
  {D.}~\bibnamefont {Loss}}, \bibinfo {author} {\bibfnamefont {M.}~\bibnamefont
  {Sherwin}}, \ and\ \bibinfo {author} {\bibfnamefont {A.}~\bibnamefont
  {Small}},\ }\href {\doibase 10.1103/PhysRevLett.83.4204} {\bibfield
  {journal} {\bibinfo  {journal} {Phys. Rev. Lett.}\ }\textbf {\bibinfo
  {volume} {83}},\ \bibinfo {pages} {4204} (\bibinfo {year}
  {1999})}\BibitemShut {NoStop}%
\bibitem [{\citenamefont {Tanamoto}\ \emph {et~al.}(2008)\citenamefont
  {Tanamoto}, \citenamefont {Maruyama}, \citenamefont {Liu}, \citenamefont
  {Hu},\ and\ \citenamefont {Nori}}]{PhysRevA.78.062313}%
  \BibitemOpen
  \bibfield  {author} {\bibinfo {author} {\bibfnamefont {T.}~\bibnamefont
  {Tanamoto}}, \bibinfo {author} {\bibfnamefont {K.}~\bibnamefont {Maruyama}},
  \bibinfo {author} {\bibfnamefont {Y.-x.}\ \bibnamefont {Liu}}, \bibinfo
  {author} {\bibfnamefont {X.}~\bibnamefont {Hu}}, \ and\ \bibinfo {author}
  {\bibfnamefont {F.}~\bibnamefont {Nori}},\ }\href {\doibase
  10.1103/PhysRevA.78.062313} {\bibfield  {journal} {\bibinfo  {journal} {Phys.
  Rev. A}\ }\textbf {\bibinfo {volume} {78}},\ \bibinfo {pages} {062313}
  (\bibinfo {year} {2008})}\BibitemShut {NoStop}%
\bibitem [{\citenamefont {Maruyama}\ and\ \citenamefont
  {Nori}(2008)}]{PhysRevA.78.022312}%
  \BibitemOpen
  \bibfield  {author} {\bibinfo {author} {\bibfnamefont {K.}~\bibnamefont
  {Maruyama}}\ and\ \bibinfo {author} {\bibfnamefont {F.}~\bibnamefont
  {Nori}},\ }\href {\doibase 10.1103/PhysRevA.78.022312} {\bibfield  {journal}
  {\bibinfo  {journal} {Phys. Rev. A}\ }\textbf {\bibinfo {volume} {78}},\
  \bibinfo {pages} {022312} (\bibinfo {year} {2008})}\BibitemShut {NoStop}%
\bibitem [{\citenamefont {Gon\ifmmode~\mbox{\c{t}}\else \c{t}\fi{}a}\ and\
  \citenamefont {van Loock}(2011)}]{PhysRevA.84.042303}%
  \BibitemOpen
  \bibfield  {author} {\bibinfo {author} {\bibfnamefont {D.}~\bibnamefont
  {Gon\ifmmode~\mbox{\c{t}}\else \c{t}\fi{}a}}\ and\ \bibinfo {author}
  {\bibfnamefont {P.}~\bibnamefont {van Loock}},\ }\href {\doibase
  10.1103/PhysRevA.84.042303} {\bibfield  {journal} {\bibinfo  {journal} {Phys.
  Rev. A}\ }\textbf {\bibinfo {volume} {84}},\ \bibinfo {pages} {042303}
  (\bibinfo {year} {2011})}\BibitemShut {NoStop}%
\bibitem [{\citenamefont {Gon\ifmmode~\mbox{\c{t}}\else \c{t}\fi{}a}\ and\
  \citenamefont {van Loock}(2012)}]{PhysRevA.86.052312}%
  \BibitemOpen
  \bibfield  {author} {\bibinfo {author} {\bibfnamefont {D.}~\bibnamefont
  {Gon\ifmmode~\mbox{\c{t}}\else \c{t}\fi{}a}}\ and\ \bibinfo {author}
  {\bibfnamefont {P.}~\bibnamefont {van Loock}},\ }\href {\doibase
  10.1103/PhysRevA.86.052312} {\bibfield  {journal} {\bibinfo  {journal} {Phys.
  Rev. A}\ }\textbf {\bibinfo {volume} {86}},\ \bibinfo {pages} {052312}
  (\bibinfo {year} {2012})}\BibitemShut {NoStop}%
\bibitem [{\citenamefont {Taylor}\ \emph {et~al.}(2005)\citenamefont {Taylor},
  \citenamefont {D\"ur}, \citenamefont {Zoller}, \citenamefont {Yacoby},
  \citenamefont {Marcus},\ and\ \citenamefont {Lukin}}]{PhysRevLett.94.236803}%
  \BibitemOpen
  \bibfield  {author} {\bibinfo {author} {\bibfnamefont {J.~M.}\ \bibnamefont
  {Taylor}}, \bibinfo {author} {\bibfnamefont {W.}~\bibnamefont {D\"ur}},
  \bibinfo {author} {\bibfnamefont {P.}~\bibnamefont {Zoller}}, \bibinfo
  {author} {\bibfnamefont {A.}~\bibnamefont {Yacoby}}, \bibinfo {author}
  {\bibfnamefont {C.~M.}\ \bibnamefont {Marcus}}, \ and\ \bibinfo {author}
  {\bibfnamefont {M.~D.}\ \bibnamefont {Lukin}},\ }\href {\doibase
  10.1103/PhysRevLett.94.236803} {\bibfield  {journal} {\bibinfo  {journal}
  {Phys. Rev. Lett.}\ }\textbf {\bibinfo {volume} {94}},\ \bibinfo {pages}
  {236803} (\bibinfo {year} {2005})}\BibitemShut {NoStop}%
\bibitem [{\citenamefont {Bennett}\ \emph
  {et~al.}(1996{\natexlab{b}})\citenamefont {Bennett}, \citenamefont
  {DiVincenzo}, \citenamefont {Smolin},\ and\ \citenamefont
  {Wootters}}]{PhysRevA.54.3824}%
  \BibitemOpen
  \bibfield  {author} {\bibinfo {author} {\bibfnamefont {C.~H.}\ \bibnamefont
  {Bennett}}, \bibinfo {author} {\bibfnamefont {D.~P.}\ \bibnamefont
  {DiVincenzo}}, \bibinfo {author} {\bibfnamefont {J.~A.}\ \bibnamefont
  {Smolin}}, \ and\ \bibinfo {author} {\bibfnamefont {W.~K.}\ \bibnamefont
  {Wootters}},\ }\href {\doibase 10.1103/PhysRevA.54.3824} {\bibfield
  {journal} {\bibinfo  {journal} {Phys. Rev. A}\ }\textbf {\bibinfo {volume}
  {54}},\ \bibinfo {pages} {3824} (\bibinfo {year}
  {1996}{\natexlab{b}})}\BibitemShut {NoStop}%
\bibitem [{\citenamefont {Werner}(1989)}]{PhysRevA.40.4277}%
  \BibitemOpen
  \bibfield  {author} {\bibinfo {author} {\bibfnamefont {R.~F.}\ \bibnamefont
  {Werner}},\ }\href {\doibase 10.1103/PhysRevA.40.4277} {\bibfield  {journal}
  {\bibinfo  {journal} {Phys. Rev. A}\ }\textbf {\bibinfo {volume} {40}},\
  \bibinfo {pages} {4277} (\bibinfo {year} {1989})}\BibitemShut {NoStop}%
\bibitem [{\citenamefont {Burkard}\ \emph {et~al.}(1999)\citenamefont
  {Burkard}, \citenamefont {Loss},\ and\ \citenamefont
  {DiVincenzo}}]{PhysRevB.59.2070}%
  \BibitemOpen
  \bibfield  {author} {\bibinfo {author} {\bibfnamefont {G.}~\bibnamefont
  {Burkard}}, \bibinfo {author} {\bibfnamefont {D.}~\bibnamefont {Loss}}, \
  and\ \bibinfo {author} {\bibfnamefont {D.~P.}\ \bibnamefont {DiVincenzo}},\
  }\href {\doibase 10.1103/PhysRevB.59.2070} {\bibfield  {journal} {\bibinfo
  {journal} {Phys. Rev. B}\ }\textbf {\bibinfo {volume} {59}},\ \bibinfo
  {pages} {2070} (\bibinfo {year} {1999})}\BibitemShut {NoStop}%
\bibitem [{\citenamefont {Barenco}\ \emph {et~al.}(1995)\citenamefont
  {Barenco}, \citenamefont {Bennett}, \citenamefont {Cleve}, \citenamefont
  {DiVincenzo}, \citenamefont {Margolus}, \citenamefont {Shor}, \citenamefont
  {Sleator}, \citenamefont {Smolin},\ and\ \citenamefont
  {Weinfurter}}]{PhysRevA.52.3457}%
  \BibitemOpen
  \bibfield  {author} {\bibinfo {author} {\bibfnamefont {A.}~\bibnamefont
  {Barenco}}, \bibinfo {author} {\bibfnamefont {C.~H.}\ \bibnamefont
  {Bennett}}, \bibinfo {author} {\bibfnamefont {R.}~\bibnamefont {Cleve}},
  \bibinfo {author} {\bibfnamefont {D.~P.}\ \bibnamefont {DiVincenzo}},
  \bibinfo {author} {\bibfnamefont {N.}~\bibnamefont {Margolus}}, \bibinfo
  {author} {\bibfnamefont {P.}~\bibnamefont {Shor}}, \bibinfo {author}
  {\bibfnamefont {T.}~\bibnamefont {Sleator}}, \bibinfo {author} {\bibfnamefont
  {J.~A.}\ \bibnamefont {Smolin}}, \ and\ \bibinfo {author} {\bibfnamefont
  {H.}~\bibnamefont {Weinfurter}},\ }\href {\doibase 10.1103/PhysRevA.52.3457}
  {\bibfield  {journal} {\bibinfo  {journal} {Phys. Rev. A}\ }\textbf {\bibinfo
  {volume} {52}},\ \bibinfo {pages} {3457} (\bibinfo {year}
  {1995})}\BibitemShut {NoStop}%
\bibitem [{\citenamefont {Nielsen}\ and\ \citenamefont
  {Chuang}(2000)}]{nielsen00}%
  \BibitemOpen
  \bibfield  {author} {\bibinfo {author} {\bibfnamefont {M.~A.}\ \bibnamefont
  {Nielsen}}\ and\ \bibinfo {author} {\bibfnamefont {I.~L.}\ \bibnamefont
  {Chuang}},\ }\href@noop {} {\emph {\bibinfo {title} {Quantum Computation and
  Quantum Information}}}\ (\bibinfo  {publisher} {Cambridge University Press},\
  \bibinfo {year} {2000})\BibitemShut {NoStop}%
\bibitem [{\citenamefont {Petta}\ \emph {et~al.}(2005)\citenamefont {Petta},
  \citenamefont {Johnson}, \citenamefont {Taylor}, \citenamefont {Laird},
  \citenamefont {Yacoby}, \citenamefont {Lukin}, \citenamefont {Marcus},
  \citenamefont {Hanson},\ and\ \citenamefont {Gossard}}]{Petta30092005}%
  \BibitemOpen
  \bibfield  {author} {\bibinfo {author} {\bibfnamefont {J.~R.}\ \bibnamefont
  {Petta}}, \bibinfo {author} {\bibfnamefont {A.~C.}\ \bibnamefont {Johnson}},
  \bibinfo {author} {\bibfnamefont {J.~M.}\ \bibnamefont {Taylor}}, \bibinfo
  {author} {\bibfnamefont {E.~A.}\ \bibnamefont {Laird}}, \bibinfo {author}
  {\bibfnamefont {A.}~\bibnamefont {Yacoby}}, \bibinfo {author} {\bibfnamefont
  {M.~D.}\ \bibnamefont {Lukin}}, \bibinfo {author} {\bibfnamefont {C.~M.}\
  \bibnamefont {Marcus}}, \bibinfo {author} {\bibfnamefont {M.~P.}\
  \bibnamefont {Hanson}}, \ and\ \bibinfo {author} {\bibfnamefont {A.~C.}\
  \bibnamefont {Gossard}},\ }\href {\doibase 10.1126/science.1116955}
  {\bibfield  {journal} {\bibinfo  {journal} {Science}\ }\textbf {\bibinfo
  {volume} {309}},\ \bibinfo {pages} {2180} (\bibinfo {year} {2005})},\ \Eprint
  {http://arxiv.org/abs/http://www.sciencemag.org/content/309/5744/2180.full.pdf}
  {http://www.sciencemag.org/content/309/5744/2180.full.pdf} \BibitemShut
  {NoStop}%
\bibitem [{\citenamefont {Nowack}\ \emph {et~al.}(2007)\citenamefont {Nowack},
  \citenamefont {Koppens}, \citenamefont {Nazarov},\ and\ \citenamefont
  {Vandersypen}}]{Nowack30112007}%
  \BibitemOpen
  \bibfield  {author} {\bibinfo {author} {\bibfnamefont {K.~C.}\ \bibnamefont
  {Nowack}}, \bibinfo {author} {\bibfnamefont {F.~H.~L.}\ \bibnamefont
  {Koppens}}, \bibinfo {author} {\bibfnamefont {Y.~V.}\ \bibnamefont
  {Nazarov}}, \ and\ \bibinfo {author} {\bibfnamefont {L.~M.~K.}\ \bibnamefont
  {Vandersypen}},\ }\href {\doibase 10.1126/science.1148092} {\bibfield
  {journal} {\bibinfo  {journal} {Science}\ }\textbf {\bibinfo {volume}
  {318}},\ \bibinfo {pages} {1430} (\bibinfo {year} {2007})},\ \Eprint
  {http://arxiv.org/abs/http://www.sciencemag.org/content/318/5855/1430.full.pdf}
  {http://www.sciencemag.org/content/318/5855/1430.full.pdf} \BibitemShut
  {NoStop}%
\end{thebibliography}%

\end{document}